\documentstyle [12pt] {article}

\newcommand{\gtwid}{\mathrel{\raise.3ex\hbox{$>$\kern-.75em\lower1ex
\hbox{$\sim$}}}}
\newcommand{\ltwid}{\mathrel{\raise.3ex\hbox{$<$\kern-.75em\lower1ex
\hbox{$\sim$}}}}
\newcommand{\be}{\begin{equation}}
\newcommand{\ee}{\end{equation}} \newcommand{\beqs}{\begin{eqnarray}}
\newcommand{\eeqs}{\end{eqnarray}} \def\({\left (} \def\){\right )}

\begin{document}

\begin{titlepage}

\begin{flushright}
\begin{tabular}{l}
ITP-SB-96-08    \\  February, 1996 
\end{tabular}
\end{flushright}

\vspace{8mm}
\begin{center}

{\LARGE\bf The Four Dimensional \vspace{2mm} Green-Schwarz Mechanism 
and Anomaly \vspace{1mm} Cancellation Conditions}

\vspace{4mm}
\vspace{16mm} Francisco Gonzalez-Rey \footnote{email:
glezrey@insti.physics.sunysb.edu}

\vspace{6mm} Institute for Theoretical Physics  \\ State University of
New York       \\ Stony Brook, N. Y. 11794-3840  \\

\vspace{20mm}

{\bf Abstract}
\end{center}

We consider a theory with gauge group $G \times U(1)_A$
containing: i) an abelian factor for which the chiral matter content
of the theory is anomalous $\sum_{f} q^f_A \neq 0 \neq \sum_{f} (q^f_A)^3$
; ii) a nonanomalous factor $G$. In these models, the calculation
of consistent gauge anomalies usually found in the literature as a 
solution to the Zumino-Stora descent equations is reconsidered. Another 
solution of the descent equations that differs on the terms involving 
mixed gauge anomalies is presented on this paper. The origin of 
their difference is analysed, and using Fujikawa's formalism the 
second result is argued to be the divergence of the usual chiral current.
Invoking topological arguments the physical equivalence of both 
solutions is explained, but only the second one can be technically 
called the consistent anomaly of a classically invariant theory. 
The first one corresponds to the addition of noninvariant local 
counterterms to the action. A consistency check of their physical
equivalence is performed by implementing the four dimensional string inspired
Green-Schwarz mechanism for both expressions. This is achieved adding slightly 
different anomaly cancelling terms to the original action, whose difference 
is precisely the local counterterms mentioned before. The complete anomaly 
free action is therefore uniquely defined, and the resulting constraints on 
the spectrum of fermion charges are the same. The Lorentz invariance of the 
fermion measure in four dimensions
forces the Lorentz variation of the Green-Schwarz terms 
to cancel by itself, producing an additional constraint usually overlooked
in the literature. This often happens when a dual description of the theory 
is used without including all local counterterms.

\vspace{35mm}

\end{titlepage}
\newpage
\setcounter{page}{1}
\pagestyle{plain}
\pagenumbering{arabic} \renewcommand{\thefootnote}{\arabic{footnote}}
\setcounter{footnote}{0}

\section{Introduction}

 The issue of anomaly cancellation is one of crucial importance in
field theory, since the presence of nonzero anomalies breaks the gauge
invariance of the quantum theory. This fact is best understood when
the anomaly is seen as a Jacobian factor, arising from a gauge
transformation of the path integral measure factor corresponding to
the chiral fermions of an otherwise classically (Lagrangian) gauge
invariant theory. The powerful path integral formalism requires
regularization of the anomaly term, since it is in principle an ill
defined expression \cite{fuji}. The issues of the regularization
independence of the method and related problems have been studied by various
authors. We will not attempt to discuss these problems, but merely derive
the expression for the anomaly that can also be obtained from
diagrammatic calculation. 

 A different approach, more algebraic, was developed in reference 
\cite{zumino-wess} using the properties of the gauge group. Regarding the 
anomaly as the result of a gauge variation in the effective action, 
its functional form must
be consistent with the fact that the commutator of two gauge
transformations can be reproduced by a gauge transformation involving
the structure constants of the group. In this approach, local 
functionals in gauge fields and their derivatives obeying this
condition are considered to be acceptable expressions
of the anomaly density . A method of obtaining
solutions for the consistency condition was developed in reference
\cite{zumino-zee}, but this solution is not always unique. In the case
we are considering it is only defined up to the gauge variation of
local functionals of gauge fields and their derivatives. Such
arbitrariness is usually justified by arguing that the effective action
has certain degree of indetermination since it can only be perturbatively 
defined up to such local functionals. The fact is that different anomaly 
expressions correspond to different classical actions, obtained by adding 
these functionals to a classically invariant action to make it noninvariant.
Presumably we will obtain these terms as higher order string loop effects,
while the gauge invariant action corresponds to the lowest order low energy
limit of the string theory.
The corresponding redefined effective field theories are
considered to be physically equivalent, and each author uses the
anomaly expression of his choice, very often forgetting that the action is no 
longer the original invariant one.

 As long as the theory is considered an effective one that point of
view is acceptable. However it is very common to find computation of 
the anomalies of a classically invariant theory using the solutions of
the consistency conditions. In this case the consistent anomaly can only come
from the transformation of the path integral measure, and it is
unambiguously determined. This is actually the only piece in the gauge 
variation of any effective action that cannot be removed by adding local 
counterterms to it, and this is often thought to imply that such 
expression of the anomaly is the only physically meaningful one. We will 
determine in this paper which solution of
the descent equations corresponds to the invariant Lagrangian.
The fact that this invariant classical action can be 
understood to be physically equivalent to the ones corresponding to other
solutions of the descent equations, will be analysed 
from a topological point of view. We will show that there is a common
topological number associated to the fermion measure anomaly.

 A necessary condition of the physical equivalence of these theories is 
that implementation of the Green-Schwarz mechanism must lead to the same
conditions in the spectrum of chiral fermion charges. The mechanism 
mentioned involves completing the original action with gauge noninvariant 
terms that cancel its gauge variation. The final complete action is therefore 
always the same, and the cancellation conditions should not change.

 In the Standard Model anomalies are not problematic. The sum of the 
$SU(3) \times SU(2) \times
U(1)_Y$ anomaly coefficients over all the chiral matter content of the
theory adds up to zero, and it is believed that any extension
including additional irreducible chiral representations of the gauge
group must have charges that balance the anomaly coefficients to zero. 

 However, in recent years the possibility of having extra abelian gauge
symmetries, obtained along with the SM gauge group after
compactification of a 10-dimensional superstring model with a large
gauge group $SO(32)$ or $E_8 \times E_8$ and additional background
Wilson lines on a 6-dimensional manifold, has been considered in
different  models. Quite generally \cite{casas} the low energy field 
theory limit of the compactified  superstring has a chiral matter content 
that makes one of these abelian symmetries anomalous. It is also a common 
feature in 4-dimensional free-fermionic string constructions with a 
factorized gauge group containing abelian factors \cite{kwai}. Since 
the underlying string theory is anomaly free it is believed that a 
consistent effective field theory has to be also anomaly free. 

 It is important therefore to have an accurate computation of the
posible anomalies of a theory to study their cancellation. We will 
review now the field theory calculation of the anomaly expression 
associated with the gauge group $G \times U(1)_A$. 

 In section two we give our working definition of the fermion measure
anomaly. In section three different solutions of the descent equations are
presented, for $G$ abelian and nonabelian. In section four, one of them 
is determined to be the fermion measure anomaly using the chiral version
of Fujikawa's regularization. In section five we give a topological 
interpretation of the physical equivalence of the theories corresponding 
to different solutions. In section six we study the Green-Schwarz 
anomaly cancellation mechanism, putting some emphasis in the rigorous 
derivation of the constraints imposed on the spectrum of anomalous
chiral fermion charges. In section seven we transform the theory into
a dual version common in the literature, to explain why one of the 
cancellation conditions is usually overlooked. We also examine some 
consequences of the presence of an anomalous $U(1)_A$ that seem to 
support the cancellation condition mentioned above.

\section{Definition of the anomaly}

 We must define the anomaly in both the Fujikawa and the Wess-Zumino 
approach, to understand the difference between them.

 If we perform a local gauge tranformation  with parameter
$\alpha(x)=\alpha^a(x) T^a$ on the fields of a Yang-Mills  theory
coupled to chiral fermions in the fundamental representation of $G\times
U(1)_A$ (so that $T^a=\lambda^a$ antihermitian for a $G$ transformation and
$T^a=q_A$ for a $U(1)_A$ transformation), using the fact that these
fermions are chiral we can rewrite their gauge transformation as a
chiral transformation with the charge containing the chirality sign

\beqs 
\psi'_L &=& \exp(- \alpha(x)) \psi_L = \exp(- \alpha(x)
 \gamma_5 \gamma_5)  \psi_L = \exp (- \alpha(x) \gamma_5) \psi_L \\
 \psi'_R &=& \exp(- \alpha(x)) \psi_R = \exp(- \alpha(x) \gamma_5
 \gamma_5)  \psi_R = \exp (+ \alpha(x) \gamma_5) \psi_R
\eeqs

\noindent
so in general we will write the gauge transformation acting on chiral
fermions as

\be
\psi'_L = \exp(- n_{L} \alpha(x) \gamma_5) \psi ;
 \bar{\psi'_L} = \bar{\psi_L} \exp(- n_L \alpha(x) \gamma_5)
\ee

 Following the analysis of Fujikawa \cite{fuji}, we can compute the 
infinitesimal gauge  variation of the path integral of the chiral Euclidean
theory and we will find an extra term coming from the Jacobian 
of the transformation in the measure 

\beqs
& Z' = \int \prod_{x} DA'_\mu (x) D\bar{\psi'}(x) D\psi'(x) e^{ -
   S(A'_\mu,\bar{\psi'},\psi') }  \nonumber  & \\ 
&  = \int \prod_{x} DA_\mu (x) D\bar{\psi}(x) D\psi(x) \(1+\int d^4 x
  tr  \alpha (x) A(x) - \delta S(A_\mu,\bar{\psi},\psi) \) 
 e^{- S(A_\mu,\bar{\psi},\psi) } &
\eeqs

 If $\delta S = 0$, a nonvanishing anomaly $A(x)\neq 0$ makes the gauge 
invariance break at the quantum level. In this case, since the anomalous 
variation corresponds to the fermion measure only, we
can restrict our study to the gauge variation of the effective action

\be
\int D \bar{\psi} D \psi \exp \( -\int d^4 x i \bar{\psi} \not{D(A)} P_L
 \psi \) = \exp \Gamma (A) \; ; \; \delta \Gamma(A) = \int d^4 x \alpha^a D_\mu
 {\partial \Gamma(A) \over \partial A^a_\mu}
\ee

Since we can rewrite

\be
{\partial \Gamma(A) \over \partial A^a_\mu }= <-i \bar{\psi} \gamma^\mu
 \lambda^a P_L \psi> = J^{\mu a}_L 
\ee

\noindent 
in the Fujikawa approach we will identify the gauge anomaly density 
with the covariant divergence of the usual chiral gauge current 
$(D_\mu J^\mu_L)^a$. For an abelian
anomaly we will identify it with the divergence $\partial_\mu J^\mu_L$. 

 If we add local terms in gauge fields to the action, the gauge
variation of the new action will correspond to the divergence of a
different current ${\partial \Gamma(A) / \partial A^a_\mu}$, so we
should not confuse it with the previous definition.

\section{Solutions of the descent equations}
 
 We will compute now the solutions of the descent equations for our factor 
gauge group.

 In reference \cite{zumino-zee} it was shown that for a simple
nonabelian gauge group the chiral anomaly $\alpha(x) A(x)$ obtained 
by Bardeen \cite{bardeen} through diagrammatic calculation and verifying 
the Wess-Zumino consistency conditions 

\be
(\delta_{\theta^1} \delta_{\theta^2} - \delta_{\theta^2} \delta_{\theta^1})
 Z = \delta_{[\theta^1 ,\theta^2]} Z
\ee

\noindent
can be reproduced by the solution of the descent equation 

\be
\delta \omega_{2n+1} = d \omega^1_{2n}
\ee

\noindent
where $\omega_{2n+1}$ is the generalized Chern-Simons form defined by 
the conveniently normalized $2n+2$ dimensional Chern character $1/192 \pi^2 
tr F^{n+1}=d \omega_{2n+1}$. For a nonabelian simple gauge group in 
$2n=4$ dimensions this gives the well known result 

\be
tr \alpha(x) D_\mu J^\mu_L = {1 \over 24 \pi^2} \epsilon^{\mu \nu 
 \rho \sigma} tr \alpha(x) \partial_\mu (A_\nu \partial_\rho A_\sigma 
 + {1 \over 2} A_\nu A_\rho A_\sigma )
\ee

 We can try to use the descent equations for our factor gauge group
with field strength $F_{\mu \nu}=[D_\mu,D_\nu]=q^f_A B_{\mu \nu} + 
\lambda_{_G a} F^a_{\mu \nu}$, and see if we reproduce the fermion
measure anomaly. The Zumino-Stora method would start
with the following Bose symmetric Chern character in 6 dimensions 

\beqs
& tr \{ (q^f_A B_{\mu \nu} + F_{\mu \nu})(q^f_A B_{\rho \sigma} + 
 F_{\rho \sigma})(q^f_A B_{\alpha \beta} + F_{\alpha \beta}) \}
 \epsilon^{\mu \nu \rho \sigma \beta \alpha}  \nonumber & \\
 = & \( (q^f_A)^3 B_{\mu \nu}B_{\rho \sigma}B_{\alpha \beta} + tr F_{\mu \nu} 
 F_{\rho \sigma}F_{\alpha \beta} + 3 q^f_A B_{\mu \nu} tr F_{\rho \sigma}
 F_{\alpha \beta} \) \epsilon^{\mu \nu \rho \sigma \alpha \beta} &
\eeqs

The associated 5-form is easy to compute, although not unique. 
An obvious choice is \cite{baulieu},\cite{preskil} 

\beqs
\omega_5 & = {1 \over 96 \pi^2} [(q^f_A)^3 B_\beta B_{\mu \nu}
 B_{\rho \sigma} + tr \{A_\beta F_{\mu \nu} F_{\rho \sigma} - 
 {1 \over 2} A_\beta A_\mu A_\nu F_{\rho \sigma} + {1 \over 10} 
 A_\beta A_\mu A_\nu A_\rho A_\sigma \} +  \nonumber   & \\
& 3 c_1 tr \{ q^f_A B_\beta F_{\mu \nu} F_{\rho \sigma} \} + 
   3 c_2 q^f_A B_{\beta \mu} \omega^G_{_3 \nu \rho \sigma} ]
 \epsilon^{\beta \mu \nu \rho \sigma}  &
\label{chern-simons1}
\eeqs

\noindent
where $c_1+c_2=1$ and $B_\mu$ is the anomalous gauge field. Different 
choices of $c_1,c_2$ differ by a total
derivative $(\Delta c)\; d(B \omega^G_3)$. The last term includes the usual 
Chern-Simons 3-form

\be
\omega_3^G = tr \{A_\nu \partial_\rho A_\sigma + {2 \over 3} A_\nu 
 A_\rho A_\sigma \}  \epsilon^{\nu \rho \sigma}
\ee

\noindent
whose exterior derivative gives the 4-dimensional Chern character

\be
\partial_\mu \omega^G_{_3 \nu \rho \sigma} \epsilon^{\mu \nu \rho \sigma} 
 = 1/4 F_{\mu \nu}  F_{\rho \sigma} \epsilon^{\mu \nu \rho \sigma}
\ee

\noindent
and whose gauge variation is 

\be
\delta_G \omega_3^G = \partial_\nu tr \alpha_G (x) \partial_\rho A_\sigma
\ee

 We can see that this 6-dimensional Chern character does not uniquely
define the 5-form $\omega_5$ whose gauge variation yields the consistent
Wess-Zumino anomaly in four dimensions, because of the different 
possibilities for
the linear combination of the last two terms subject to the condition 
$c_1 + c_2 = 1$.

 Correspondingly the expression of the solution of the descent equations
is not unique. If this solutions were to coincide with divergence of 
the chiral currents as defined above, we would write after summation 
over all chiral fermion representations 
\cite{preskil}

\beqs
& \int \alpha_A (x) \sum_{f} n_f \partial_\mu J^\mu_{_A f} = 
 {1\over 24 \pi^2}
 \int  \alpha_A(x) \sum_{f} n_f q^f_A [{1 \over 4} (q^f_A)^2 B_{\mu \nu} 
 B_{\rho \sigma} + {3 \over 4} c_1 tr \{ \lambda_a \lambda_b \} F^a_{\mu \nu} 
 F^b_{\rho \sigma} ] \epsilon^{\mu \nu \rho \sigma}   &  \label{descabel} \\
& \int  tr \alpha_G (x) \sum_{f} n_f D_\mu J^\mu_{_G f} =  
 {1\over 24 \pi^2} \int  \alpha^a_G (x) \sum_{f} n_f [ tr \{ \lambda_a 
\lambda_b \lambda_c \} \partial_\mu (A^b_\nu \partial_\rho 
 A^c_\sigma + {f^c_{de} \over 4} A^b_\nu A^d_\rho A^e_\sigma )
   \nonumber   &  \\
& + 3/2 c_2 tr \{ q^f_A \lambda_a \lambda_b \} B_{\mu \nu} \partial_\rho  
 A^b_\sigma ] \epsilon^{\mu \nu \rho \sigma} &
\label{descentanom}
\eeqs

 It is commonly argued that since the last term in each of the equations
(\ref{descentanom}) can be obtained as the gauge variation of a local 
counterterm $B_\mu \omega^G_3$, and the effective action can only be 
defined in perturbation theory up to such local counterterms, we can 
set one of the coefficients $c_2=0$. This is an obscure statement. A
fair attitude is to regard the effective theory as equivalent to 
a noninvariant classical theory. For a nonanomalous group $G$, using
this choice of coefficients only the mixed and pure $U(1)$ anomalies 
corresponding to the abelian gauge variation survive and we obtain the 
results usually cited in the literature \cite{harvey}. However it is not
usually indicated what is the noninvariant action corresponding to 
this gauge variation.

As we will see, when additional interactions are included in the action 
to completely cancel the Wess-Zumino anomaly, the term in (\ref{descentanom}) 
balanced by a nongauge coupling is that proportional to $c_2$. According
to the authors of reference \cite{lerche},in this set of additional 
interactions only the nongauge coupling is physically relevant . From 
that point of view it seems more natural to use the mentioned freedom to
set the coefficient $c_1=0$. A third possibility favored 
by Bose symmetry in the decomposition of the Chern-Simons form 
\cite{alvarez} is $c_1=1/3,c_2=2/3$.

 On the other hand if we take the more rigid definition of the anomaly,
in the Fujikawa approach such indetermination cannot be allowed. If
the quantum theory is to be consistent, either the fermion measure 
anomalies cancel summing over the fermion content
of a theory with a classically gauge invariant action, or they cancel
against the gauge variation of some terms in a non-invariant
classical action. It can only cancel against the variation of
noninvariant terms in the action. The action must be a well defined 
object with all local interactions clearly specified for a given theory, 
and in a consistent theory the chiral anomaly cannot admit any
indetermination. Only one solution of the decent equations can give
the fermion measure anomaly. We will try to determine this below, invoking
topological arguments. 

 It is also interesting to consider the posibility that the nonanomalous 
factor is an abelian group $U(1)_{NA}$. In this case the descent
equations method would start from a Chern character including an 
additional term

\be
\( (q^f_A)^3 B_{\mu \nu}B_{\rho \sigma}B_{\alpha \beta} + (q^f_{NA})^3 
 F_{\mu \nu} F_{\rho \sigma}F_{\alpha \beta} + 3 q^f_A (q^f_{NA})^2 
 B_{\mu \nu} tr F^{NA}_{\rho \sigma} F^{NA}_{\alpha \beta} + 3 (q^f_A)^2 
 q^f_{NA} B_{\mu \nu} B_{\rho \sigma} F^{NA}_{\alpha \beta} \)
\ee

\noindent
that properly normalized can be obtained as the exterior derivative of 
the 5-form

\beqs
& \omega_5 = {1 \over 96 \pi^2} [(q^f_A)^3 B_\beta B_{\mu \nu}
 B_{\rho \sigma} + (q^f_{NA})^3 A_\beta F^{NA}_{\mu \nu} 
 F^{NA}_{\rho \sigma} + 3 c_1 tr q^f_A (q^f_{NA})^2 B_\beta 
 F^{NA}_{\mu \nu} F^{NA}_{\rho \sigma}   \nonumber &     \\
&  + 3 c_2 q^f_A (q^f_{NA})^2 A_\beta B_{\mu \nu} F^{NA}_{\rho \sigma} + 
  3 d_1 B_\beta B_{\mu \nu} F^{NA}_{\rho \sigma} + 3 d_2 A_\beta 
  B_{\mu \nu} B_{\rho \sigma} ] \epsilon^{\beta \mu \nu \rho \sigma}  &
\eeqs

\noindent
where as before $d_1+d_2=1=c_1+c_2$ and different choices of the coefficients 
are obtained by adding a total derivative to $\omega_5$. If the solution 
to the descent equations gives the gauge variation of the fermion measure,
the corresponding gauge anomaly is

\beqs
& \int  \alpha_A (x) \sum_{f} n_f \partial_\mu J^\mu_{_A f} = & \\
& {1\over 24 \pi^2} \int  \alpha_A(x) \sum_{f} n_f q^f_A 
 [{1 \over 4} (q^f_A)^2 B_{\mu \nu} B_{\rho \sigma} +
 {3 c_1 \over 4} (q^f_{NA})^2 F^{NA}_{\mu \nu} F^{NA}_{\rho \sigma} + 
 {3 d_1 \over 4} q^f_A q^f_{NA} B_{\mu \nu} F^{NA}_{\rho \sigma} ] 
  \epsilon^{\mu \nu \rho \sigma}   &  \nonumber   \\
&  & \nonumber  \\
& \int  \alpha_{NA} (x) \sum_{f} n_f \partial_\mu J^\mu_{_{NA} f} =  & 
  \label{abeldescentanom} \\
& {1\over 24 \pi^2} \int  \alpha_{NA} (x) \sum_{f} n_f q^f_{NA} 
 [ {1 \over 4} (q^f_{NA})^2 F^{NA}_{\mu \nu} F^{NA}_{\rho \sigma} 
 + {3 c_2 \over 4} q^f_A q^f_{NA} B_{\mu \nu} F^{NA}_{\rho \sigma} +
  {3 d_2 \over 4} (q^f_A)^2 B_{\mu \nu} B_{\rho \sigma} ]
 \epsilon^{\mu \nu \rho \sigma} & \nonumber
\eeqs

 From our definition of chiral currents, $J^\mu_{_{NA} f}$ can be 
obtained from $J^\mu_{_A f}$ by replacing $q^f_A \leftrightarrow 
q^f_{NA}$. Therefore the $U(1)_{NA}$ anomaly functional
must be the same as the result of replacing $\alpha_A (x) q^f_A
\leftrightarrow \alpha_{NA} (x) q^f_{NA}$ in the $U(1)_A$ anomaly
functional. This imposes $c_1=1/3=d_2,d_1=c_2$ which combined
with the constraints $c_1+c_2=1=d_1+d_2$ fixes the coefficients to be
$d_1=2/3=c_2$ as required by Bose symmetry. We see that the consistent 
Fujikawa anomaly is completely determined in this case.

 As we shall see below, it is not possible to cancel the anomaly terms 
with a coefficient $\sum_{f}  n_f (q^f_A)^2 q^f_{NA} \neq 0$ using the
Green-Schwarz mechanism. Imposing the vanishing of
such coefficient for a particular chiral content of the theory many 
authors do not include these terms in the general expression of the
solution of the descent equations for $U(1)_A \times U(1)_{NA}$, and 
just choose the coefficients
$c_1=1,c_2=0$. We will prove that although this choice is formally
incorrect if we want to write the divergence of the usual chiral current, both 
expressions correspond to the same
winding number of the Weyl determinant around a loop of gauge
transformations. If such winding number is the relevant physical
quantity both results must be considered physically
equivalent. Correspondingly we expect the anomaly cancellation
conditions resulting from the Green-Schwartz mechanism to be the same.

 For $G$ nonabelian, it turns out that none of the solutions of the
descent equations presented so far provides the fermion measure anomaly. We 
will now find the correct solution and prove it to be physically equivalent to 
({\ref{descentanom}).

 We can try a different solution for the 5-form $\tilde{\omega_5}$, 
using the general formula \cite{zumino-zee} for $2n$ dimensions

\be
\tilde{\omega}_{2n+1} = (n+1) \int dt t^n tr \{ A (dA + t A^2)^n \}_{|_{n=2}} 
 = A dA dA + 3/2 A A^2 dA + 3/5 A A^2 A^2
\ee

\noindent
after substitution of $A_\mu = q^f_A B_\mu + A^a_\mu \lambda_a$ 
for our factor gauge group. The pure $U(1)$ and $G$ terms are the same 
as in the previous version, but the mixed terms are now very different

\beqs
& \tilde{\omega}_5 = {1 \over 96 \pi^2} [ (q^f_A)^3 B_\beta B_{\mu \nu}
 B_{\rho \sigma}) + 4 
 tr \{ A_\beta \partial_\mu A_\nu \partial_\rho A_\sigma + {3 \over 2} 
 A_\beta A_\mu A_\nu \partial_\rho A_\sigma + {3 \over 5} A_\beta A_\mu 
 A_\nu A_\rho A_\sigma \}  & \nonumber \\ 
& + 4 tr q^f_A B_\beta \{ \partial_\mu A_\nu \partial_\rho 
 A_\sigma + 3/2 A_\mu A_\nu \partial_\rho A_\sigma \} 
 + 4 tr q^f_A \partial_\beta B_\mu \{ 2 A_\nu \partial_\rho A_\sigma + {3
 \over 2} A_\nu A_\rho A_\sigma \} ] \epsilon^{\beta \mu \nu \rho \sigma}  &
\label{chern-simons2}
\eeqs

 The exterior derivative of this expression gives the same Chern
character as the (\ref{chern-simons1}) but the
cross terms cannot be obtained from (\ref{chern-simons1}) for any
choice of $c_1, c_2$, and the solution of the consistency conditions
that it yields is not the same as (\ref{descentanom}). The gauge
variation of (\ref{chern-simons2}) gives a result consistent with the
substitution $A_\mu = q^f_A B_\mu + A^a_\mu \lambda_a$ in the general
solution of the consistency conditions found by \cite{zumino-zee} for
a simple gauge group

\be 
{i^n \over (2 \pi)^n (n+1) \!} n(n+1) \int_0^1 dt (1-t) str\{ \alpha (x) 
 d_x [ A (F^t)^{n-1}] \}|_{n=2} = {1 \over 24 \pi^2} \alpha(x) (A dA +
 {1 \over 2} A A^2) 
\label{graldescentanom}
\ee

\noindent
where $F^t = t d_x A + t^2 A^2$. The pure $U(1)_A$ and $G$
anomalies coincide in both solutions but the mixed anomalies are
different 

\beqs
& \int  \alpha_A (x) \sum_{f} n_f \partial_\mu J^\mu_{_A f} =  & 
  \nonumber   \\
& {1\over 24 \pi^2}
 \int  \alpha_A(x) \sum_{f} n_f q^f_A [ {1 \over 4} (q^f_A)^2 B_{\mu \nu} 
 B_{\rho \sigma} + tr \{ \lambda_a \lambda_b \} \partial_\mu (A^a_\nu 
 \partial_\rho A^b_\sigma + {f_{bde} \over 4} A^a_\nu A^d_\rho A^e_\sigma)
  ] \epsilon^{\mu \nu \rho \sigma}  & 
\label{descentanom1}
\eeqs

\noindent
for a $U(1)_A$ variation and  

\beqs
& \int  tr \alpha_G (x) \sum_{f} n_f D_\mu J^\mu_{_G f} =  
 {1\over 24 \pi^2} \int  \alpha^a_G (x) \sum_{f} n_f [ tr \{ \lambda_a 
 \lambda_b \lambda_c \} \partial_\mu (A^b_\nu \partial_\rho 
 A^c_\sigma + {f^c_{de} \over 4} A^b_\nu A^d_\rho A^e_\sigma) & \nonumber  \\
& + tr \{ \lambda_a q^f_A \lambda_b \} \partial_\mu (2 B_\nu \partial_\rho 
 A^b_\sigma + {f_{bde} \over 4} B_\nu A^d_\rho A^e_\sigma)
 ] \epsilon^{\mu \nu \rho \sigma} &
\label{descentanom2}
\eeqs

\noindent
for a $G$ variation. 

 It is easy to see that for $G=U(1)_{NA}$ the
substitution $A_\mu = q^f_A B_\mu + q^F_{NA} A^{NA}_\mu$ in the
general solution (\ref{graldescentanom}) gives the same result as the
Bose symmetric choice $d_2=1/3=c_1,c_2=2/3=d_1$ in the previous
formulation. This increases our confidence that the prescription to
obtain the solution of the descent equations that yields the divergence of
the chiral current, is to substitute
the gauge connection of the factor group in (\ref{graldescentanom}).

 For nonabelian $G$ the discrepancy found in both solutions can be 
understood by realising \cite{itabashi} 
that the most general solution to the consistency conditions should
use the following generalized Chern character 

\be
\Omega_{2n+2} = c_{n+1} tr F^{n+1} + \sum_{m} c_{n,m} tr F^m F^{n-m} +
 \sum_{m,l} c_{n,m,l} tr F^m F^{m} F^{n-m-l} + \cdots
\label{gralchernchar}
\ee

This 2n+2-form can be written as the exterior derivative of

\be
c_{n+1} \tilde{\omega}_{2n+1} + \sum_{m} c_{n,m} \tilde{\omega}_m 
 tr F^{n-m} + \cdots
\label{gralchern-sim}
\ee

\noindent
up to a closed form. In our 4-dimensional $G \times U(1)_A$ case we
notice that the the first solution of the consistency conditions is 
obtained after ignoring the first term in (\ref{gralchern-sim}) for 
the calculation of the mixed anomalies. We also notice that the second
solution of the consistency conditions only makes use of such term.
It was argued in reference \cite{petersen} that for
nonabelian groups in 2n dimensions only the term with coefficient
$c_{n+1}$ needs to be considered since its result is the one
coinciding with perturbative calculations. 

 Since for our factor gauge group the first two terms in (\ref{gralchernchar}) 
coincide if we set the coefficients equal to one, the corresponding 
exterior derivatives of the two terms in (\ref{gralchern-sim}) give the 
same result. However when we use the Zumino-Stora method to find the 
divergence of the chiral currents defined above, we should
use the first term in (\ref{gralchernchar}) as we shall argue, and 
therefore we should consider the second solution (\ref{descentanom1})-
(\ref{descentanom2}) of the consistency conditions the one providing
the fermion measure anomaly.

\section{The regularized fermion measure anomaly}

 To support our assertion that for $G$ nonabelian the second solution of 
the descent equations is the one that formally reproduces the gauge 
variation of the fermionic measure , we will 
consider now the path integral calculation of the anomaly. We will
reproduce for our factor gauge group the analysis of references 
\cite{alvarez-g} \cite{einhorn}, in which they
modify Fujikawa's path integral analysis in the case of chiral fermions.
This computation gives the same result as the diagrammatic calculation
\cite{bardeen} of the anomaly for a simple nonabelian group, and that 
fact leads to the belief that the regularization used by
\cite{alvarez-g} is correct.

 For a chiral fermionic action 

\be
i \bar{\psi} \not{D \; \;} 1/2(1+\gamma_5) \psi = \bar{\psi} \gamma^\mu
 (\partial_\mu + A_\mu) 1/2(1+\gamma_5) \psi \;\; , A_\mu = q_A^f B_\mu 
 + A_\mu^a \lambda_a
\ee

\noindent
the Fujikawa method requires extending the fermion sector to gauge
invariant righthanded partners, that only introduce a numerical factor
after their integration in the path integral. The total fermionic
action is thus 

\be
\bar{\psi} \not{\tilde{D} \; \;} \psi =  \bar{\psi} \gamma^\mu 
 \( \partial_\mu + A_\mu {1+\gamma_5 \over 2} \) \psi =  \bar{\psi} \gamma^\mu 
 \( \partial_\mu {1-\gamma_5 \over 2} + [\partial_\mu + A_\mu] {1+\gamma_5 
 \over 2} \) \psi
\ee

\noindent
where the operator $\not{\tilde{D} \; \;}$ has well defined
eigenvalues as opposed to the original $\not{D \; \;}$ that maps
positive chirality spinors to negative chirality ones. We can use these
eigenvalues to construct a Fujikawa type regulator for the anomaly.

 Working in Euclidean space-time after a Wick rotation of the action,
the authors in \cite{alvarez-g} expand the Dirac fermion operators in 
a basis of eigenstates of the modified Dirac operator (which being 
nonhermitian forces us to consider both eigenstates of $i \not{\tilde{D} \
; \;}$ and $(i \not{\tilde{D} \; \;})^\dagger$ )

\begin{eqnarray*}
\psi(x)= \sum_{n} a_n \phi_n(x) & , & \bar{\psi}(x)=\sum_n \bar{b}_n
 \chi^\dag_n(x)    \\
i\not{\tilde{D} \; \;} \phi_n(x) = k_n \phi_n(x) & , &
 (i \not{\tilde{D} \; \;})^\dag \chi_n(x) = k^*_n \chi_n(x)   
\end{eqnarray*}

\noindent
where the coefficients $a_n , \bar{b}_n$ are Grassman valued. The 
commuting
Lorentz spinors $\phi_n(x)$ and $\chi_n(x)$ obey orthonormality and 
closure relations

\be
\int d^4 x \chi^\dag_m(x) \phi_n(x) = \delta_{m,n} \; \; ; \sum_{n}
 \phi_n^{\beta,p} (x) \chi^{\dag \: \gamma,q}_n(y) = \delta^{\beta
 ,\gamma} \delta^{p,q} \delta^4(x-y)
\ee

 After a gauge transformation the transformed Grassman coefficients can 
be obtained as

\beqs
& a'_n=\int d^4 x \chi^\dag_n(x) (\psi^L{}'(x) +\psi^R(x)) = & \nonumber  \\
& =  \int d^4 x \sum_{m} \( a_m \chi^\dag_n (x) \exp(- n_L \alpha(x) 
 \gamma_5)  P_L \phi_m (x) + a_m \chi^\dag_n (x) P_R \phi_m (x) \) & \nonumber
   \\
& = \sum_{m} (C^L_{n,m} + \delta^R_{n,m}) a_m  &
\eeqs

\noindent
and similarly for the transformed $\bar{b}_n$

 Rewriting the path integral fermionic measure in terms of the grassman 
coefficients it is very simple to find the Jacobian

\be
\prod_{x} D \bar{\psi^L}'(x) D \psi^L{}'(x) =\prod_{m} d\bar{b}'_m 
 \prod_{n} da'_n = (\det \bar{C})^{-1} \prod_{m} d\bar{b}_m  (\det C)^{-1})
 \prod_n da_n
\ee

 For an infinitesimal transformatiom we can evaluate $\det C =\exp(tr
\log C)$ by expanding $C$ to first order in $\alpha$
so the jacobian can also be written to first order in the transformation
parameter

\be  
(\det C)^{-1} = \exp( tr \int d^4 x \phi^{L \dag}_m (x)
 n_L \alpha(x) \gamma_5 P_L \phi_n (x) )\simeq 1 + \sum_{n} 
 \int d^4 x  \chi^\dag_n(x)  n_L \alpha(x) \gamma_5 P_L \phi_n (x) 
\ee

 Similarly the we find the Jacobian resulting from the transformation of 
the Dirac conjugate fermion and combining both

\beqs
& (\det C^R)^{-1} (\det C^L)^{-1} = \exp\{ tr \int d^4 x \chi^\dag_m (x) 
 n_L \alpha(x) \gamma_5 \phi_n (x) &  \nonumber   \\
\simeq & 1 + \sum_{n} \int d^4 x \chi^\dag_n(x) n_L \alpha(x) 
 \gamma_5 \phi_n (x) &
\eeqs

 The closure relations that the basis $\phi_n , \chi_n$ obeys reduce the
anomaly  of the lefthanded fermion field to

\be 
\int i d^4 x \alpha(x) A(x) = \int d^4 x \alpha^a(x) n_ftr \{ T_a
 \gamma_5 \delta^4(x-x) \}
\ee

 Although $tr \gamma_5 = 0$  the anomaly as it stands is an ill-defined 
quantity because of the divergence introduced by the delta function. 
To regularize \'a la Fujikawa we need to include a weight factor 
$\exp((-i \not{\tilde{D} \; \;)^2} / M^2)$ that reduces to one in the limit 
$M \rightarrow \infty$. With this regularization, we expand the
eigenstates of the modified Dirac operator on plane waves 

\beqs  
\alpha (x) A (x) = & \lim_{x \rightarrow y , M \rightarrow \infty} & \sum_{n}
 \phi^\dagger_n(y) g_a \alpha^a(x) T_a \gamma_5 \exp \(-(i\not{\tilde{D} \; 
 \;)^2}/M^2 \) \phi_n(x) =   \nonumber   \\
 = & \lim_{x \rightarrow y, M \rightarrow \infty} & \alpha^a(x) \int {d^4 k
 \over (2\pi)^4} e^{-iky} tr T_a \gamma_5 \exp \(-(i \not{\tilde{D} \; \;)^2}
 /M^2 \) e^{ikx}
\eeqs

 Rewriting the operator $(i \not{\tilde{D} \; \;)^2}$ as in reference
\cite{alvarez-g} 

\be
(i \not{\tilde{D} \; \;)^2} = \(i( \not{\partial P_R} + \not{D 
 P_L})\)^2 = - (\not{\partial \;\;} \not{D \; \;} P_L + \not{D \; \;}
\not{\partial \; \;} P_R )
\ee

\noindent
and decomposing the chirality operator $\gamma_5 = P_L - P_R$ we can 
separate the regularized anomaly

\be
tr T_a \gamma_5 \exp \(-(i \not{\tilde{D} \; \;)^2} /M^2 \) = 
 tr T_a P_L \exp (\not{\partial \;\;} \not{D \; \;} /M^2) - tr T_a P_R \exp
 (\not{D \; \;} \not{\partial \; \;} /M^2)
\ee

 Pulling the plane wave factor to the left, we expand the exponential around 
$-k^\mu k_\mu$ in both terms. After integration on $k$ the 
only abnormal parity terms that survive in the limit $M \rightarrow
\infty$ are those proportional to $M^{-4}$. The result of adding the
contributions from both terms is \cite{alvarez-g} the consistent anomaly

\be
{1 \over 24 \pi^2} \int d^4 x tr n_L \{ \alpha (x) 
  \partial_\mu (A_\nu \partial_\rho A_\sigma + {1 \over 2}
 A_\nu A_\rho A_\sigma) \} \epsilon^{\mu \nu \rho \sigma} 
\label{anom}
\ee

 After substituting $A_\mu = q_A^f B_\mu + A_\mu^a \lambda_a$  and $\alpha
= \alpha_A + \alpha_a \lambda_a$, we obtain the gauge anomaly by 
summing the contributions of all lefthanded ($n_L = +1$) and
righthanded ($n_R = -1$) fermion representations of $G \times U(1)_A$. 
 This is precisely the second solution 
(\ref{descentanom1})-(\ref{descentanom2}) of the descent equations.

\section{Topological equivalence of different solutions of the descent 
equations}

 Now that we have found a regularization for the anomalous variation
of the fermion measure that supports the alternative solution of the
descent equations presented
in this paper, further study of the connection between the Zumino-Stora 
method and the consistent anomaly is needed to understand why both solutions 
are physically equivalent. 

 For that purpose, it is necessary to explain why a solution of the
descent equations yields the same result as the local density resulting 
from the gauge variation of the path integral fermionic measure, for a 
chiral theory in $2n$ dimensions. This mysterious coincidence has been
explained by Alvarez-Gaum\'e and Ginsparg by noticing that the anomaly
functional can be understood as an infinitesimal gauge variation in 
the phase of the Weyl determinant that formally defines the fermionic 
effective action (the norm of such determinant is gauge invariant)

\be
\int D \bar{\psi} D \psi \exp (\int dx i \bar{\psi} \not{\tilde{D}(A)\;}
 \psi) = \exp \Gamma (A) = \det i \not{\tilde{D}(A)\;} = 
 |\det i \not{\tilde{D}(A)\;}| \; e^{i w(A,\theta)}
\ee

 Using topological arguments, the integrated gauge variation of the Weyl 
determinant,i.e. its winding number, is proved to be equivalent to the
index of the Dirac operator \cite{alvarez-g}
in a 2n+2-dimensional space constructed by tensoring the 2n-dimensional 
Euclidean space (compactified to a 2n-sphere) with a disc parametrized 
by polar coordinates $t$ and $\theta$. In this extended space gauge 
connections are defined as 

\beqs
& A(t,\theta)_\mu = t A^{\theta}_\mu = t g^{-1}(\theta,x) (A_\mu + 
 \partial_\mu) g(\theta,x) & \\
& A_\theta = t g^{-1}(\theta,x) \partial_\theta g(\theta,x) ,\;  A_t = 0  & \\
& g(0,x)=1=g(2 \pi,x) & 
\eeqs

\noindent
so that the coordinate $\theta$ parametrizes a path of gauge
transformed connections on the border of the disc. Infinitesimal
displacements $\theta \rightarrow \theta +\delta \theta$ along the 
path can be seen as additional gauge transformations with gauge 
parameter $\alpha = g^{-1} d_\theta g$.

 When we integrate the gauge variation of the effective action over a
loop of gauge transformed configurations we find the winding number for
the phase of the Weyl determinant, which coincides with the index of
the Dirac operator on the disc

\be
\int^{2 \pi}_0 d \theta \delta_\theta \Gamma(A^\theta) = \int^{2 \pi}_0 
 d \theta {\partial w(A, \theta) \over \partial \theta} 
 = ind ( i \not{D_{2n+2}})_{|_{S^{2n} \times Disc }}
\ee

 To apply the index theorem on this 2n+2-dimensional space bounded by
the edge of the disk, boundary conditions are implicitly given by
mapping the disk to the upper patch of a 2n+2-sphere and defining
gauge connections on a lower patch covering the rest of the sphere. 
Parametrizing the distance to the pole on each patch with $0<t<1$ 
and $0<s<1$ the boundary between lower and upper patches $t=1=s$ is 
identified with the edge of the disk, and the gauge connection is 
defined on this lower patch as

\beqs
& A(s,\theta)_\mu = A^{\theta}_\mu  & \\
& A_\theta = 0 = A_s   & 
\eeqs

\noindent
so that the transition function on the boundary between patches is
precisely $g(\theta,x)$.
The corresponding extended field strength is then constructed on the
upper patch

\beqs
& {\cal F} = (d_x + d_t + d_\theta) {\cal A}(t,\theta) + {\cal A}^2(t,\theta) =
 F_x(t,\theta) + t d_x \alpha + A^\theta dt + \alpha dt + t d_\theta
 A^\theta   &  \\
& F_x(t,\theta) = t d_x A^\theta + t^2 (A^\theta)^2 \; , 
 F_{t,\mu} = A^\theta_\mu \;, F_{\theta \mu} =
 (t^2-t) [\alpha, A_\mu^\theta ]+ \; , F_{t \theta} = \alpha & 
\eeqs

\noindent
and the lower patch 

\be
{\cal F} = (d_x + d_s + d_\theta) {\cal A}(s,\theta) + {\cal A}^2(s,\theta) =
 F_x
\ee

 Using the Atiyah-Singer index theorem we can reproduce the index of
the Dirac operator by integrating the Chern character constructed from 
this generalized field strength over the (2n+2)-dimensional space. Since
the Chern character is an exact 2n+2-form, the result of this
integration is only the difference between the Chern-Simons 2n+1-forms of the
lower and upper patches evaluated at the boundary $t=1=s$:

\beqs
& ind (i \not{D_{2n+2}}) = {i^{n+1} \over (2 \pi)^{n+1} (n+1)!}
 \int_{S^2 \times S^{2n}} tr {\cal F}^{n+1} = & \nonumber   \\
& {i^{n+1} \over (2 \pi)^{n+1} 
 (n+1)!} \int_{S^1 \times S^{2n}} d \theta d^{2n} x \( \omega_{2n+1} (t=1)
 - \omega_{2n+1} (s=1) \) &
\eeqs

 Since the Chern-Simons form on the lower patch does not contain any
$d \theta$ or $ds$ its contribution vanishes. On the upper
patch we need only to consider the component of the Chern-Simons form
with no $t$ lower index. This component is linear in $\alpha(x)$ because
the form can only contain one $d \theta$ differential. The gauge 
parameter and the factor it multiplies can be written as a $\theta$ 
exterior derivative, precisely the gauge variation of the Chern-Simons 
form as prescribed by the Zumino-Stora method

\be
\omega_{2n+1} = \omega^0_{2n+1} + \alpha d\theta \wedge \omega^1_{2n}
 \; , \; \delta \omega_{2n+1} = d \omega^1_{2n}
\ee

 We arrive at two conclusions from this analysis. First, when we apply the 
Zumino-Stora method to find the divergence of the chiral currents defined 
above, i.e. the fermion measure anomaly, 
we should only consider the first term in (\ref{gralchernchar}). 
As a plus we get the normalization factor of the nonabelian anomaly
from the index theorem normalization constant \cite{alvarez-g}.

 The second conclusion is that we have found a characterization
of physical equivalence for the theories we consider. Two solutions 
of the descent equations 
are physically equivalent if the corresponding Chern-Simons 5-forms 
belong to the same cohomology class. When we integrate these forms over 
$S^1 \times S^{2n}$ we find the same result. As we have seen this
amounts to integrate the infinitesimal gauge transformation of the
action over a loop of such transformations. The local counterterms
that make the Lagrangian noninvariant do not contribute to this loop,
and all the theories in the same class have the same integrated
anomaly. It is always equal to the winding number of the Weyl
determinant, i.e. the integrated fermion measure anomaly.
 
 If two Chern-Simons 5-forms differ by a closed form,  when we
integrate over the parameter $\theta$ we will not in general find the 
same winding number for the phase of the Weyl determinant, and we
cannot consider the corresponding expressions for the anomaly density 
physically equivalent. Different Weyl determinant winding numbers correspond 
to different fermion measure anomalies, but if the expression of this
consistent anomaly is truly regularization independent, this situation
is not possible for a consistent theory.

 In the case we are studying, the difference between both versions of
the Chern-simons form can be straightforwardly computed

\beqs
& \tilde{\omega_5} - \omega_5  = B tr ( dA dA + {3 \over 2} A^2 dA) + dB
 tr (2 dA dA + {3 \over 2} A A^2) - 3 c_1 B tr ( dA dA + 2 A^2 dA )  & 
 \nonumber     \\
& -  3 c_2 dB tr (A dA + {2 \over 3} A A^2) =  (1- 3 c_1) B tr (dA dA) + 
 ({3 \over 2} - 6 c_1) B tr (A^2 dA) + & \nonumber   \\
&  (2- 3 c_2) dB tr (A dA) + ({3 \over 2} - 2 c_2) dB tr (A A^2) & 
\eeqs

 Terms with two derivatives can only be obtained from $tr(B A dA)$ because
$tr(dB A^2)=0$, while terms with one derivative can only come from
differentiating $tr B A A^2$. It is easy to see that only $c_1+c_2=1$
can simultaneously obey $1-3c_1=-(2-3c_2)$ and $3/2-6c_1 =-3(3/2-2c_2)$, 
converting the expression above into a total derivative. 

 Therefore, we learn that although the second prescription for the
consistent gauge 
anomaly is the one that we would find by proper regularization of the Fujikawa 
method, the first solution of the descent equations, while representing 
a different 
density does not change the global winding number of the weyl
determinant for any choice of the coefficients $c_1+c_2=1$. 
It seems that the relevant physical information carried by the anomaly 
is contained in this topological
quantity.

 It is worth remarking that this justification of the use of the
descent equations only works for a very particular choice in reference
\cite{alvarez-g2} of the extended gauge connections on the two patches
of the 2n+2-dimensional space. A different but also acceptable choice
in references \cite{alvarez-g} and \cite{alvarez} leads directly to
the general solution (\ref{graldescentanom}) and does not allow to
make contact with the gauge variation of the Chern-Simons 2n+1-forms 
as prescribed by the descent equations.

\section{The Green-Schwarz mechanism and anomaly cancellation conditions}

 We have mentioned that the fermion measure anomaly is the only piece
in the gauge variation of the effective action that cannot be balanced
by the gauge variation of local terms in gauge fields. If it does not
vanish when summing its coefficient over all chiral fermion
representations, the theory is in principle unacceptable as a quantum
theory. Nevertheless, in certain cases when the Chern character
factorizes, it is possible to balance this nonzero anomaly against the
special gauge variation of a bosonic degree of freedom coupled to
gauge fields \cite{green-schwarz} and some local counterterms. This is 
known as the Green-Schwarz
mechanism. Once the fermion measure anomaly is cancelled by such local
terms, any additional piece in the gauge variation of the effective
action can be cancelled by the gauge transformation of gauge field
counterterms as we know. Therefore even if we use different solutions 
of the descent equations as anomaly densities of physically equivalent 
actions, the constraint on the physical
spectrum of particles that the cancellation imposes must be the same.

 To test the equivalence of different anomaly expressions, let us study 
the anomaly cancellation conditions derived from implementing the
4-dimensional Green-Schwarz mechanism. For completeness, we will consider
the possibility of having mixed gravitational-$U(1)_A$ anomalies.
We would compute them from a similar method, but using the product
$(-1/8) Tr F tr R R$ in the descent equations instead \cite{alvarez-g2}. Pure 
gravitational anomalies do not exist in
four dimensions but for $4n+2$ dimensions they would be computed from 
$tr R R R$ using a generalization of the arguments that justify the
use of the descent equations in the pure gauge case \cite
{alvarez-g2}. The mixed gravitational-$U(1)_A$ anomaly arising from a
$U(1)_A$ variation of the path integral measure is uniquely determined

\be
- {1\over 24 \pi^2} \int  \alpha_A(x) \sum_{f} n_f q^f_A 
 {1 \over 8 \times 4} 
 tr R_{\mu \nu} R_{\rho \sigma} \epsilon^{\mu \nu \rho \sigma}
\ee

 Assuming that the gauge group $G$ is anomaly free $ \sum_{f} tr \lambda_a
\lambda_b \lambda_c = 0$, complete cancellation of the anomaly
(\ref{descentanom}) through the four dimensional 
Green-Schwarz mechanism, requires the mixed gauge, pure $U(1)_A$
gauge and the mixed gravitational-$U(1)_A$ anomalies to balance the 
gauge variation of the adittional couplings introduced in the field 
theory action

\be
S \rightarrow S - {1\over 24 \pi^2} \int  ({c_A \Lambda \over 2 g_A} 
 M_{\mu \nu} B_{\rho \sigma} + 
 {c_G 3 c_1 \over g_A g^2_G} B_\mu \omega^G_{_3 \nu \rho \sigma} -
 {c_L \over g_A g^2_L} 
 B_\mu \omega^L_{_3 \nu \rho \sigma}) \epsilon^{\mu \nu \rho \sigma}
\label{green-schwarz1}
\ee

 The gauge fields are normalized in this couplings so that they do not 
contain the coupling constant implicit in the covariant derivative. 
Consistent with this normalization the spin connection is also formally
normalized by a coupling constant to be defined later by analogy with
the gauge coupling constants, although it does not have the same
physical meaning. 

 The field $M$ is the bosonic, second rank tensor in the supergravity
multiplet that couples to $tr F^n$ in $2n$ dimensions. Such coupling
is a one loop string effective term \cite{lerche} not present in the
tree level action derived from the low energy limit of the string.
We have also the mass parameter $\Lambda=\sqrt{2}/ \kappa$ defined from the 
field strength of $B$ \cite{green-schwarz}, for our factor gauge
group 

\be
H = dB + {\kappa \over \sqrt{2}} ({\omega^A_3 \over g^2_A} + 
 {\omega^G_3 \over g^2_G} -{\omega^L_3 \over g^2_L})
\label{hfield}
\ee
 
 By analogy with 10-dimensional supergravity \cite{chapline}, this 
field strength is defined to include Chern-Simons gauge forms
normalized by the gauge coupling constants. To achieve anomaly
cancellation we need to extend the field strength to include also 
Lorentz Chern-Simons forms coming from a higher order string effective term
\cite{green-schwarz}. It is common to use the same form of
normalization as the one appearing in the gauge forms. In 10-dimensional
string inspired theories, since the anomaly free gauge groups considered 
have a unique coupling constant, such normalization is well defined 
\cite{green-schwarz}. In the 4-dimensional analog different gauge factors
have different gauge couplings at tree level given by 
$1/g^2_G=k_G /g^2$ \cite{ginsparg} where $k_G$ is the Kac level
of $G$ and $g^2$ is the v.e.v. of the dilaton. It is not clear what 
normalization should be used for the Lorentz Chern-Simoms form unless 
the coefficient of this effective term is computed from the 4-dimensional
string theory considered. We will formally define
a ``Kac level'' for gravity $1/g^2_L=k_L/g^2$. It is only a
normalization factor defining the gravitation term in (\ref{hfield}),
in principle having no relation with any Kac-Moody algebra.

Gauge invariance of the field strength $H$ imposes a nontrivial gauge 
(and Lorentz) variation of the antisymmetric supergravity tensor

\be 
\delta M_{\mu \nu} = ({1 \over \Lambda  g^2_A} \alpha_A \partial_\mu 
 B_\nu + {1 \over \Lambda  g^2_G} tr \{ \alpha_G \partial_\mu A_\nu \}
 + {1 \over \Lambda  g^2_L} tr \{ \Delta_L \partial_\mu 
 \omega^L_\nu \} )\epsilon^{\mu \nu}
\ee 

\noindent
where $\omega^L_\mu$ is the spin connection.

 The cancellation of the anomaly generated by a $U(1)$ transformation 
fixes the coefficients of the added couplings to be 

\be
c_A = g^3_A \sum_{f} n_f (q^f_A)^3 \; ; \; c_G = g_A g^2_G \sum_{f} n_f 
 tr q^f_A \lambda_a \lambda_b \; ; \; c_L = {g_A g^2_L \over 8} \sum_{f} 
 n_f q^f_A 
\label{chargeconstraint}
\ee

 The last sum runs over all chiral fermions
with anomalous charges since all of them couple to gravity.

Cancellation of the mixed anomaly generated by a $G$ transformation is
possible only if 

\be
{c_A \over 2 g_A g^2_G}  - {3 c_1 c_G \over 2 g_A g^2_G} = 3/2 c_2
\sum_{f} n_f tr q^f_A \lambda_a \lambda_b 
\ee

 The additional condition $c_1+c_2=1$ imposes a nontrivial constraint on 
the anomalous charges of the chiral fermion content of the theory 

\be
g^2_A \sum_{f} n_f (q^f_A)^3 = 3 g^2_G \sum_{f} n_f tr q^f_A \lambda_a
 \lambda_b
\ee

 Similarly, the absence of any anomaly under Lorentz transformations
imposes

\be
g^2_A \sum_{f} n_f (q^f_A)^3 = {g^2_L \over 8} \sum_{f} n_f q^f_A 
\label{lorentzcancel}
\ee

The cancellation condition we have found from the Lorentz transformation
is not the usual one found in the literature . The
reason for this is that some of the anomaly cancelling terms are 
usually transformed to a dual version of the theory, in which they are
Lorentz invariant, while the rest of the terms are forgotten. It is
the omitted terms the ones that impose this last condition. We will
come back to this point later when the dual version is studied.

An argument supporting the cancellation conditions (\ref{lorentzcancel})
is the coefficient of
the $M dB$ coupling obtained from the string theory 1-loop amplitude
\cite{lerche} , which seems to be 

\be
g^{-2}_A / 48 \pi^2 \sum_{f} n_f q^f_A
\label{mcoeff} 
\ee
 
 Unless $\sum_{f} n_f q^f_A = 2 \sum_{f} n_f (q^f_A)^3$, which is
precisely our condition (\ref{lorentzcancel}) when $k_L=4k_A$, we will not 
have the right coefficient for the $M dB$ term. This relation between 
anomalous charges differs
by a factor of four from the actual relation found in the particular 
model of reference \cite{casas}. It might be due to a different 
normalization convention of the field $H$ used by \cite{lerche} in 
their computation of (\ref{mcoeff}).
 
 Now we must make sure it is possible to remove the anomalies in our 
second solution of the descent equations (\ref{descentanom1})-
(\ref{descentanom2})
when $\sum_{f} n_f tr \{ \lambda_a q^f_A \lambda_b \} \neq 0
\neq \sum_{f} q^f_A q^f_A q^f_A $, by a 4-dimensional Green-Schwarz
mechanism similar to that discussed for the first solution of the 
descent equations. The cancellation conditions should be the same
in both cases.
 
 Following the path we tried before, we can try adding to the
classical action the coupling of the supergravity second rank tensor
with the anomalous field strength, and the coupling of the anomalous
gauge field with the three form that defines the mixed $U(1)_A$
anomaly. 
(since the mixed gravitational-$U(1)_A$ anomaly has not
changed, the cancellation term and resulting condition are the same as
before)

\be
S \rightarrow S - {1\over 24 \pi^2} \int  ({c_A \over g_A}
 \Lambda M_{\mu \nu} \partial_\rho B_\sigma 
 + {c_G \over g_A g^2_G} B_\mu \tilde{\omega}^G_{_3 \nu \rho \sigma} -
 {c_L \over g_A g^2_L} B_\mu \omega^L_{_3 \nu \rho \sigma} )
 \epsilon^{\mu \nu \rho \sigma}
\label{green-sch}
\ee

 Since the second solution of the descent equations is the one
corresponding to the fermion measure anomaly, the Green-Schwarz terms
that we are adding are precisely the effective couplings that complete
the gauge invariant action to make an anomaly free theory.

 The 3-form

\be
\tilde{\omega}^G_3 = tr \{A_\nu \partial_\rho 
 A_\sigma + {1 \over 2} A_\nu A_\rho A_\sigma \} \epsilon^{\nu \rho \sigma} 
 = (A^a_\nu \partial_\rho A^a_\sigma + {1 \over 4} f_{ade} A^a_\nu 
 A^d_\rho A^e_\sigma) \epsilon^{\nu \rho \sigma}
\ee

\noindent
has the following $G$ gauge variation 

\be
(2 \partial_\nu \alpha^a (x) \partial_\rho A^a_\sigma - {f_{ade} \over 4} 
 \partial_\nu \alpha^a (x) A^d_\rho A^e_\sigma) \epsilon^{\nu \rho \sigma} 
\ee

 It is easy to see that after partial integration, the $U(1)_A$ and $G$ 
gauge variations of (\ref{green-sch}) can cancel the gauge and mixed
gravitational-gauge anomalies we found provided

\beqs
& g^3_A \sum_{f} n_f q^f_A q^f_A q^f_A = c_A & \\
& g_A g^2_G \sum_{f} n_f tr \{q^f_A \lambda_a \lambda_b \} = c_G & \\
& g_A g^2_G \sum_{f} n_f tr \{ \lambda_a q^f_A \lambda_b \} = c_A - 2c_G &
\eeqs

 The nontrivial constraint on the spectrum of fermion anomalous 
charges $3 c_G = c_A$ arises again as expected. The vanishing Lorentz 
variation of (\ref{green-sch}) imposes the same constraint as before.

 Using the string tree level definition of the gauge coupling constants
an interesting consequence of the cancellation conditions appears  
if the nonanomalous factor $G$ is a factor group itself 
$G=G_1 \times G_2$. The constraints we have found impose that 
the sum of anomalous charges for the spectrum of fermions transforming
under each subfactor is proportional to the corresponding Kac
level \cite {ibanez}

\be
\sum_{f} n_f q^f_A tr \lambda^{G_1}_a \lambda^{G_1}_b / \sum_{f} n_f (q^f_A)^3 
 = k_a / 3 k_A
\label{ibanezcond}
\ee

 As promised before we will perform the analysis of the Green-Schwarz
mechanism for the case $G=U(1)_{NA}$. We could try to include the
following additional anomaly cancelling terms in the action

\be
{1\over 24 \pi^2} \int  ({c_{NA} \over g_{NA}} \Lambda M_{\mu \nu} 
 \partial_\rho A_\sigma + {c_G \over g_{NA} g^2_A} A_\mu tr 
 \tilde{\omega}^A_{_3 \nu \rho \sigma}) \epsilon^{\mu \nu \rho \sigma}
\label{green-sch-abe}
\ee

\noindent
but the first term introduces a non vanishing $U(1)_{NA}$ variation
without a corresponding pure $U(1)_{NA}$ anomaly to cancel, therefore the
coefficient $c_{NA}$ must be zero. It is easy to see that the second
term cannot cancel by itself the mixed anomalies in
(\ref{abeldescentanom}). Another nontrivial constraint arises again 

\be
\sum_{f} n_f (q^f_A)^2 q^f_{NA} = 0
\ee
 
 This result agrees with the cancellation conditions found in the
literature for such mixed abelian anomalies.

\section{Dual version of the Green-Schwarz terms and axion-like couplings}

 As a final exercise, we will now review how to rewrite the Green-Schwarz 
terms in the action so that a coupling of a pseudoscalar to the 
4-dimensional Chern character appears in the effective field theory. 
First we consider the case when the anomaly is given by the factorized 
solutions of the descent equations \cite{harvey},
and then we will 
reproduce the analysis for the second form of the anomaly presented 
in this paper, showing that the same pseudoscalar coupling is found.

 Since the antisymmetric tensor field $M_{\mu \nu}$ only contains one
degree of freedom \cite{lopez}, it is common in the literature to replace 
it by a pseudoscalar through a duality transformation.
Again following reference \cite{harvey}, which provides the most
elegant explanation, we can see that the field $M_{\mu \nu}$ appears
on the kinetic term $H{}^*H$ containing its gauge and Lorentz invariant 
field strength (\ref{hfield}) that obeys a generalized Bianchi identity

\be
d H = {1 \over 4 \Lambda} ({F^2_A \over g^2_A} + {tr F^2_G \over g^2_G} 
 - {tr R^2 \over g^2_L})
\ee

\noindent
and it also appears in the Green-Schwarz term ${c_A \Lambda \over g_A} 
M d B^A$. The field equation of $M_{\mu \nu}$ 

\be
- \partial_\rho H^{\rho \mu \nu} + {c_A \Lambda \over 48 \pi^2 g_A} 
 \partial_\rho B_\tau \epsilon^{ \mu \nu \rho \tau} = 0
\label{mfieldeqn}
\ee

\noindent
allows us to write the field srength in terms of the divergence
of a pseudoscalar for $M_{\mu \nu}$ on shell

\be
-H^{\rho \mu \nu} + {c_A \Lambda \over 48 \pi^2 g_A} B_\tau 
 \epsilon^{\mu \nu \rho \tau} = \partial_\tau \theta (x) 
 \epsilon^{\nu \mu \rho \tau}
\label{mshell}
\ee

 This relation provides an alternative description of theory when we
use it to rewrite the kinetic term $[{}^* d \theta - 1/(48 \pi^2 g_A) \;c_A
\Lambda {}^*B][d \theta - 1/(48 \pi^2 g_A)\;c_A \Lambda B]$ and the first of
the Green-Schwarz terms

\be
{c_A \Lambda \over 24 \pi^2 g_A} \int M_{\mu \nu} \partial_\rho 
 B_\sigma \epsilon^{\mu \nu \rho \sigma} = {c_A \Lambda \over 24 \pi^2 g_A} 
 \int M_{\mu \nu} {48 \pi^2 g_A \over c_A \Lambda} \partial_{\rho} 
 H^{\rho \mu \nu} 
\ee

\noindent
partial integrating we find

\beqs 
& -2 \int \partial_\rho M_{\mu \nu}  H^{\rho \mu \nu} = 2 \int \( 
 H_{\rho \mu \nu} - {1 \over \Lambda} (g^{-2}_A \omega^A_3 + g^{-2}_G 
 \omega^G_3 - g^{-2}_L \omega^L_3)_{\rho \mu \nu} \) H^{\rho \mu \nu} = &  
  \nonumber   \\
& 2 \int \( - H_{\rho \mu \nu} H^{\rho \mu \nu} + {1 \over \Lambda}
 (g^{-2}_A \omega^A_3 + g^{-2}_G \omega^G_3 - g^{-2}_L 
 \omega^L_3)_{\rho \mu \nu} (\partial_\tau \theta (x) - 
 {c_A \Lambda \over 48 \pi^2 g_A} B_\tau) \epsilon^{\tau \rho \mu \nu} \) &
\eeqs

 The first term combines with the kinetic term, while the second one
and the additional piece in the original anomaly cancelling term
(\ref{green-schwarz1}) provide the new cancellation terms in this
description of the theory 

\beqs
& 2 \int [ - {\theta (x) \over \Lambda} \partial_\tau (g^{-2}_A \omega^A_3 + 
 g^{-2}_G \omega^G_3 - g^{-2}_L \omega^L_3)_{\rho \mu \nu} - 
 {1 \over 24 \pi^2 g_A} c_A B_\tau (g^{-2}_G \omega^G_3 - g^{-2}_L 
 \omega^L_3)_{\rho \mu \nu} +    &   \nonumber    \\
& {1 \over 24 \pi^2 g_A} B_\tau (3 c_1 c_G g^{-2}_G \omega^G_3 - c_L g^{-2}_L
 \omega^L_3)_{\tau \mu \nu} ] \epsilon^ {\tau \rho \mu \nu}    &
\label{dualgreen-schwarz}
\eeqs

\noindent
where the antisimetrization of Lorentz indices has made the term 
$\omega^A_3 B$ vanish. The first three terms give the axion-like 
couplings of the pseudoscalar

\be
2 \int {\theta (x) \over g^2 \Lambda} 1/4 (k_A tr B_{\mu \nu} B_{\rho \sigma} 
 + k_G tr F_{\mu \nu} F_{\rho \sigma} - k_L tr R_{\mu \nu} R_{\rho \sigma} ) 
 \epsilon^{\mu \nu \rho \sigma}
\label{axioncoupling}
\ee

 Now the equation of motion of the pseudoscalar contains the Bianchi
identity of the field strength $H$, while the equation of motion of
the tensor field $M$ (\ref{mfieldeqn}) can be seen as a Bianchi
identity for $\theta(x)$ \cite{harvey}. This is therefore a dual
description of the theory. 

 To see how the anomaly cancellation happens in the dual model we
notice that the pseudoscalar must shift under a $U(1)_A$ tranformation to 
mantain the gauge invariance of $H$ in (\ref{mshell})

\be
\delta_A \theta (x) = {c_A \Lambda \over 48 \pi^2 g_A} \alpha_A (x) 
\label{axiondm}
\ee

 This gauge shift under $U(1)_A$ is precisely the one needed for the 
terms in (\ref{dualgreen-schwarz}) to cancel the anomalous terms in 
(\ref{descabel}). Under a $G$ gauge variation we obtain anomaly
cancellation provided the condition $c_A-3c_1c_G=3c_2c_G$ is met again.

 In this formulation, the duality transformation apparently simplifies 
the theory when $c_1=1,c_2=0$. In that case the cancelling terms are 
just the axion-like couplings,
all the anomalies are generated by $U(1)_A$, and the gauge tranformation 
(\ref{axiondm}) achieves their complete cancellation. However, we have
already seen that such a choice amounts to include compensating local 
counterterms together with the minimal action, that make the total action
noninvariant. It is common in the literature 
to make the choice of coefficients mentioned and forget the additional
terms in the action. Without them the minimal action plus dualised
axion-like Green-Schwarz terms are invariant
under Lorentz transformations, and the nontrivial constraint
(\ref{lorentzcancel}) is never found. 

 The cancellation condition involving $\sum_f n_f q^f_A$ that can be 
usually found in the literature, assumes that the last term in 
(\ref{axioncoupling}) is the one responsible for the cancellation of 
the mixed $U(1)_A$-gravitational anomaly. In that case we would find 
a constraint similar to (\ref{ibanezcond}), i.e. the ratio of the 
gravitational ``Kac'' level and the anomaly coefficient $\sum_f n_f q^f_A$
should be commensurate to the corresponding ratios for the gauge factors.

\be
{3 \sum_f n_f q^f_A tr \lambda_a \lambda_b \over k_a} = {\sum_f n_f 
 (q^f_A)^3 \over K_A} = {\sum_f n_f q^f_A \over 8 k_L}
\ee

 We can see that including all the Green-Schwarz terms as we should,
the $U(1)_A$ gauge variation of the second term in (\ref{dualgreen-schwarz})
balances the variation of the axion-like couplings $\theta F^2$ and 
$\theta R^2$ in the first term,
so that the anomaly cancellation is provided by the third term. Therefore
we recover the correct cancellation condition (\ref{chargeconstraint}) 
by keeping all the Green-Schwarz terms.

 Keeping these additional local counterterms in the dual version of the 
action, we can still identify the pseudoscalar $\theta / g^2 \Lambda$,
with the imaginary part of
the scalar component of the chiral superfield that defines the gauge
coupling. In supersymmetric theories this coupling is given by the
dilaton v.e.v. multiplying the gauge kinetic function (up to a global
normalization factor)

\beqs
& f_{ab}(S) W^a_\alpha W^{b\alpha} \mid_{\theta \theta} = k_a (S+S^\dag)
 \delta_{ab} F^a_{\mu \nu} F^{^b \mu \nu} + k_a (S-S^\dag) \delta_{ab}
 F^a_{\mu \nu} \tilde{F}^{^b \mu \nu} \simeq   &  \nonumber   \\
& {1 \over g_a^2} F_{_a \mu \nu} F_{_a}^{\mu \nu} + k_a {\theta (x) 
 \over \Lambda'}F_{_a \mu \nu} \tilde{F}_a^{\mu \nu} &
\eeqs  
 
 From our identification, the mass scale $\Lambda'=g^2/ \Lambda$.
The axion-like coupling $\theta tr R \tilde{R}$ is not present in minimal
supersymmetric theories, but it is usually understood as a higher order 
string effect.

 Performing the same duality transformation on the Green-Schwarz terms
that cancel the fermion measure anomaly
(\ref{descentanom2}), we can see that we obtain
again axion-like couplings. In this case we have an explicit expression 
of the complete anomaly free action, and the noninvariant terms 
cannot be accidentally omitted. 

 The field equation of $M$ and the definition of the pseudoscalar are the
same as before. Rewriting the Green-Schwarz interactions (\ref{green-sch}) 
in terms of such pseudoscalar we find 

\beqs
& {1\over 24 \pi^2} \int  ({c_A \Lambda \over g_A} M_{\mu \nu} 
 \partial_\rho B_\sigma 
 + {c_G \over g_A g^2_G} B_\mu tr \tilde{\omega}^G_{_3 \nu \rho \sigma}) 
 \epsilon^{\mu \nu \rho \sigma} = -2 \int H_{\rho \mu \nu} 
 H^{\rho \mu \nu} + \int [ - 2 {\theta (x) \over g^2 \Lambda} 
 \partial_\tau (k_A \omega^A_3 + &  \nonumber  \\
& + k_G \omega^G_3 - k_L \omega^L_3)_{\rho \mu \nu} + {1\over 24 \pi^2 g_A} 
 B_\tau (c_G g^{-2}_G \tilde{\omega}^G - c_A g^{-2}_G \omega^G_3 + (c_A-c_L) 
 g^{-2}_L \omega^L_3)_{\rho \mu \nu} ] \epsilon^{\mu \nu \rho \sigma}   &
\eeqs

 We find again the coupling of the pseudoscalar to the Pontryagin 
densities 

\be
\theta(x) / (g^2 \Lambda) (k_A tr F^2_A + k_G tr F_G^2 - 
k_L tr R^2)
\ee

\noindent
that allows to identify the dual degree of freedom with the
axion partner of the dilaton. The anomaly cancellation conditions are
of course the same as before.

 An interesting phenomenological consequence of this identification is
the appearance of a Fayet-Iliopoulos term corresponding to the
anomalous gauge group \cite{dine}. The $U(1)_A$ gauge transformation
of the axion forces us to include the corresponding vector supermultiplet 
in the K\"ahler potential of the dilaton superfield to mantain its gauge
invariance

\be
\ln ( S + S^\dag) \rightarrow \ln ( S + S^\dag + {c_A \Lambda \over 48 \pi^2 
 \Lambda'})
\ee

 This term contributes to the Euler-Lagrange equation of the auxiliary 
field $D$ in the $U(1)_A$ vector multiplet

\be
D_A = {c_A \over 48 \pi^2 \kappa^2} + \sum_{f} q_f \mid \chi_f
 \mid^2 = 0 
\ee

\noindent
where $\chi_f$ are the scalar partners of chiral fermions.

In order to avoid supersymmetry
breaking by the anomalous D term at the unification scale, some $U(1)_A$
charged scalars must develop a v.e.v. that breaks the anomalous symmetry.
If they correspond to flat directions of the superpotential so that 
supersymmetry is preserved, this mechanism provides interesting 
additional symmetry breaking that can reduce some of the large gauge 
groups resulting from compactification. It has also been used to explain
the hierarchy of effective Yukawa couplings \cite{ross} \cite{robert}
\cite{ramond}.

The presence of the constant in the effective D term has been proved 
\cite{ichenose} by determining a nonzero string 1-loop amplitude
that can be identified with it. The 
coefficient found however, is proportional to $\sum_f n_f q^f_A$ 
instead of $c_A$. This seems to hint again that our cancellation condition 
(\ref{lorentzcancel}) is right.

\section{Summary}

In summary, we have analysed in this paper the topological equivalence 
of different solutions of the Zumino-Stora descent equations, 
that are computed from a Chern-Simons 5-form defined up to an exterior 
derivative. Such solutions correspond to the same winding number of the 
Weyl determinant. Defining the anomaly as the gauge variation of 
the fermionic measure in the path integral, only one of these solutions
can be properly called anomaly density. The physically 
equivalence of these functionals when used as anomaly densities has been 
tested, by studying the conditions on the spectrum of anomalous charges   
that we obtain when we impose the 4-dimensional Green-Schwarz mechanism.
As expected, the conditions are the same. One of such conditions is not
the usual one found in the literature, because we have been careful to
include in the effective action all the Green-Schwarz terms that 
make the quantum theory truly gauge and Lorentz invariant.

\section*{Acknowledgments}

We thank Prof. R. Shrock for encouragement and useful suggestions. We also 
thank Prof. M. Roc\v{e}k for enlightening discussions. This research was 
partially supported by the NSF grant PHY-93-09888.

\end{document}